\documentclass[12pt]{article}
\usepackage{graphicx}
\usepackage{natbib} 
\usepackage{url} 
\usepackage{booktabs}

\usepackage{url}    
\usepackage{amsfonts}
\usepackage{bbm}
\usepackage{float}
\usepackage{amsmath}
\usepackage{xcolor}

\usepackage{algpseudocode}

\newcommand{\blind}{0}

\begin{document}

\def\spacingset#1{\renewcommand{\baselinestretch}%
{#1}\small\normalsize} \spacingset{1}


\if0\blind
{
\title{Functional-Ordinal Canonical Correlation Analysis With Application to Data from Optical Sensors}
  \author{Giulia Patanè$^{*,\dag}$,
    Federica Nicolussi$^\dag$, Alexander Krauth$^\mathsection$,\\
    G\"unter Gauglitz$^\mathsection$,
    Bianca Maria Colosimo$^\ddag$,\\ Luca Dede'$^\dag$, and
    Alessandra Menafoglio$^\dag$\\[5mm]
    $^*$ {\tt giulia.patane@polimi.it}\\[3mm]
    $\dag$ MOX, Department of Mathematics, \\ Politecnico di Milano,\\ Piazza Leonardo da Vinci 32, 20133 Milano, Italy.\\[3mm]
    $^\mathsection$ Institute of Physical and Theoretical Chemistry (IPTC), \\
Eberhard Karls Universität Tübingen, \\ Auf der Morgenstelle 18, 
 72076 Tübingen, Germany\\[3mm]
    $\ddag$ Department of Mechanical Engineering, \\ Politecnico di Milano,\\ Piazza Leonardo da Vinci 32, 20133 Milano, Italy.\\
    }
  \maketitle
} \fi

\if1\blind
{
  \bigskip
  \begin{center}
    {\LARGE\bf Functional-Ordinal Canonical Correlation Analysis With Application to Data from Optical Sensors}
\end{center}
  \medskip
} \fi

\bigskip
\begin{abstract}
\noindent
{We address the problem of predicting a target ordinal variable based on observable features consisting of functional profiles. This problem is crucial, especially in decision-making driven by sensor systems, when the goal is to assess an ordinal variable such as the degree of deterioration, quality level, or risk stage of a process, starting from functional data observed via sensors. We purposely introduce a novel approach called functional-ordinal Canonical Correlation Analysis (foCCA), which is based on a functional data analysis approach. FoCCA allows for dimensionality reduction of observable features while maximizing their ability to differentiate between consecutive levels of an ordinal target variable. Unlike existing methods for supervised learning from functional data, foCCA fully incorporates the ordinal nature of the target variable. This enables the model to capture and represent the relative dissimilarities between consecutive levels of the ordinal target, while also explaining these differences through the functional features. Extensive simulations demonstrate that foCCA outperforms current state-of-the-art methods in terms of prediction accuracy in the reduced feature space. A case study involving the prediction of antigen concentration levels from optical biosensor signals further confirms the superior performance of foCCA, showcasing both improved predictive power and enhanced interpretability compared to competing approaches.}
\end{abstract}
\noindent%
\section{Introduction}


\label{sec:intro}


%

\begin{figure}[H]
    \centering
    \includegraphics[width=1\linewidth]{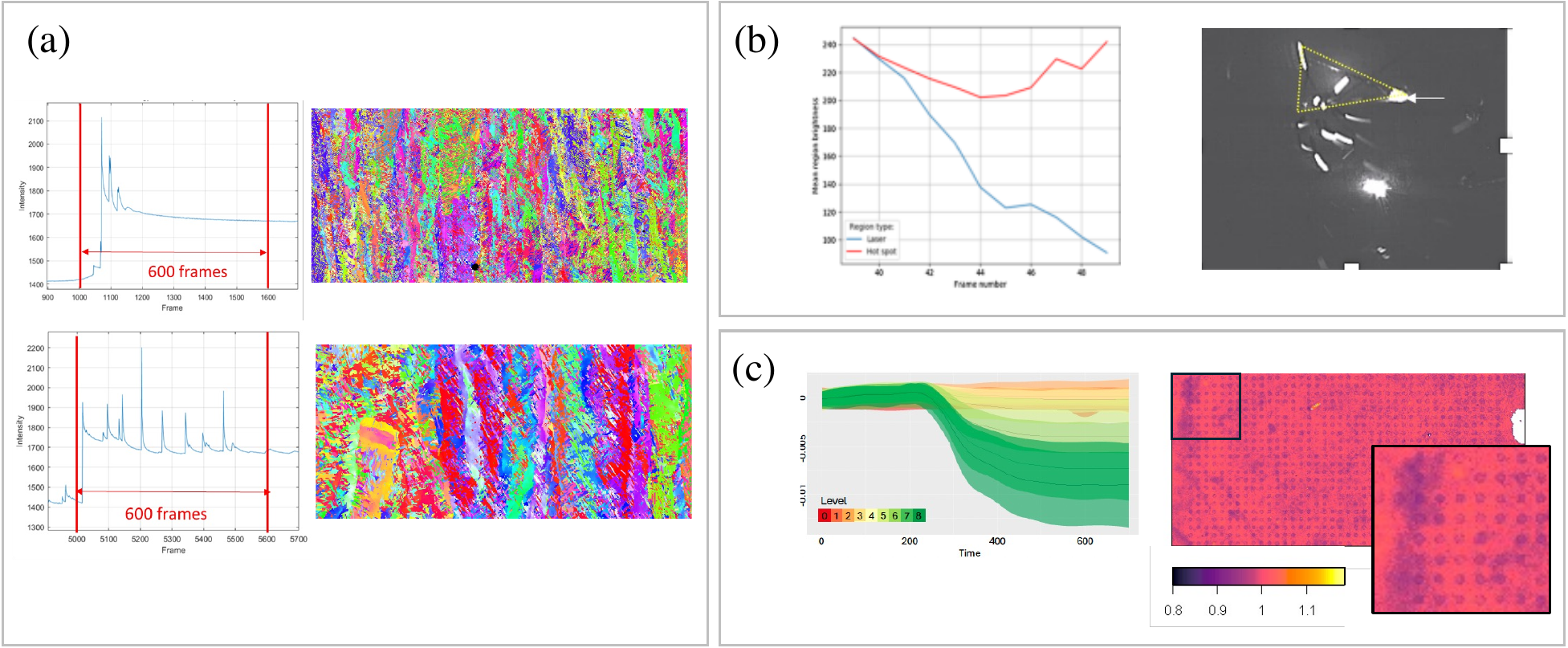}
    \caption{a) Functional data representing the cooling history observed during additive manufacturing in one location and the final microstructure (fine microstructure on the top and coarse microstructure on the bottom) b) Functional data representing two different cooling history observed during additive manufacturing with negligible or severe hot-spot phenomena (classified depending on severity) c) Functional data observed at different levels of the antibody-antigene interactions (case study described in Section \ref{motivating}).}
    \label{examples_intro}
\end{figure}

In the current industrial landscape, quality features - such as product flaws, process yield, or degradation states—are increasingly linked to data acquired via sensors (e.g., signals, images, and video images). In many cases, the quality level is measured on an ordinal scale, while the observed signal data can be modeled as functional data. This problem naturally arises in the presence of modern sensoring technologies that generate high-dimensional temporal sequences of data through video or image records. Examples can be found in manufacturing, where processes involving local thermal histories (e.g. welding, additive manufacturing, casting) often correlate final local flaws (e.g., porosity) or microstructure classifications to the spatio-temporal temperature patterns observed at the same location - see Figure image a) and b) \citep{bugatti22,yan22}. Similarly, in life-science and biotechnology applications, as the motivating example we will focus on, the reaction class observed at a given location can be linked to the local intensity patterns captured through video imaging (Figure \ref{examples_intro}, image c)).
Further examples of works that employ time-varying signals for ordinal prediction are typically found in the fields of psychometry and neuroscience \citep{psychometry,neuroscience}, medicine \citep{medical}, and again in manufacturing \citep{manufacturing,engineering}. 

In this context, it is essential to define a method that leverages functional data as a predictive driver to estimate the final quality class on an ordinal scale. Although critical for the applications, to the best of our knowledge, no available methodology leverages high-dimensional or functional signals to predict an ordinal variable. In fact, state-of-the-art methods to address a similar problem typically include a preliminary dimensionality reduction step, followed by a correlation analysis. The preliminary step aims to capture the variability of the functional data without accounting for the need to explore their ability to predict the ordinal target variable. For example, within the framework of functional regression models, several authors have suggested using Functional Principal Component Analysis (fPCA, \cite{ramsaysilvermanbook}) as a dimensionality reduction step \citep{muller05}. Here, fPCA is applied to reduce the dimensionality of the data, and then the resulting scores are used in place of functional covariates \citep{fun-pred}, functional responses \citep{pred-fun}, or both \citep{fun-fun}.
Unfortunately, when dealing with high-dimensional data, as in our case study, the global variability of the dataset can hide the ordinal patterns, significantly limiting the prediction power of the fPC scores. In the classification setting, functional Fisher's Discriminant Analysis has been proposed \citep{hastie95,ramsaysilvermanbook}, as an alternative to fPCA. However, when applied to ordinal variables, Fisher's approach may struggle to identify components that effectively discriminate between consecutive levels, especially if the ordinal levels are not evenly spaced.

Our novel contribution aims at bridging this gap by providing a statistical methodology for ordinal prediction based on a functional signal. Our novel approach jointly tackles the limitations of the current literature outlined above by proposing a novel method called functional-ordinal Canonical Correlation Analysis (foCCA). Our approach is inspired by functional Canonical Correlation Analysis \citep{ramsaysilvermanbook}, but reinterprets this in an ordinal setting, by adapting the key concept of \textit{optimal scoring} discussed by \cite{hastie95}. In our newly developed solution, foCCA allows one to reduce the dimensionality of a functional dataset while maximizing correlation of the scores with the ordinal variable of interest, thus establishing interpretable differences between adjacent levels through an appropriate weight function. We shall show that foCCA enables one to (i) effectively reduce the data complexity, (ii) perform supervised classification of the ordinal levels in the reduced space, and (iii) obtain a set of functional and ordinal directions (which play the same role as the loadings in fPCA), which can be used to enhance the model interpretation. 

In Section \ref{class_met} we formulate the problem and we recall the dimensionality reduction methods typically employed in the functional setting. Section \ref{foCCA_met} introduces foCCA, providing some insights in the discretization of the problem, the interpretation of the canonical directions and the penalty parameter selection. In Section \ref{sim_study} we show an extensive Monte Carlo simulation used to analyse the performance of foCCA and state-of-the-art methods under varied conditions. This work, though widely applicable, is primarily motivated by a monitoring challenge from biosensor data, specifically focused on assessing antigen concentration levels within specific Regions of Interest (ROIs) based on video signals that track the temporal progression of an antibody-antigen reaction. In Section \ref{motivating} we present the motivating case study and we show the performance of foCCA on its dataset. Section \ref{sec:conc} presents the main conclusions of the work. 

\section{Problem formulation and state of the art}
\label{class_met}
Let $\mathcal{C}$ be an ordinal random variable with levels from 0 to $n_\mathcal{C}$.

Suppose that we are provided with $n$ independent realizations of the functional variable $X$ and $\mathcal{C}$, then our samples consists of pairs $\{x_i(t),c_i\}_{i=1}^{n}$. The main objective is to be able to predict the level $\mathcal{C}$, knowing the functional variable $X$. For any $c\in \{1,..., n_\mathcal{C}\}$, let $D_c$ be defined as $D_c = \mathbbm{1}_{C\geq c}(\mathcal{C})$, and let $\{e_c\}_c$ be the canonical basis of $\mathbbm{R}^{n_\mathcal{C}}$.
 
Only in foCCA, we encode $\mathcal{C}$ into the random variable $Y$, defined as follow: 
\begin{equation}
    Y=\sum_{c=1}^{n_{\mathcal{C}}}{D_c e_c}
    \label{defY}
\end{equation}
Notice that, by construction, the vector $Y$ and the variable $\mathcal{C}$ are in one-to-one correspondence, hence they contain exactly the same information. 

In Section \ref{fPCA} and \ref{foFD}, we provide an overview of classical methodologies that are widely employed for dimensionality reduction, aiming to capture essential features within high-dimensional datasets in a predictive context. Understanding these classical approaches is fundamental for contextualizing and contrasting with the innovative method introduced later in this study.

\subsection{Functional Principal Component Analysis}
\label{fPCA}
PCA stands as a cornerstone in the field of dimensionality reduction, finding widespread applications across various domains, from machine learning to statistics. The PCA objective is to identify linear combinations of features that maximize the variance within a given dataset. While the introduction of PCA traces back to its roots in multivariate data analysis (\cite{hotelling33}), its scope has considerably expanded in recent years to encompass diverse contexts and various data types. This adaptability has allowed PCA to be applied to entities ranging from points and functions to images and beyond. In this broad context, fPCA (and its smoothed versions) allows to understand and extract patterns from functional observations. In the following, we set the notation and briefly recall the smoothed version of the methodology; we refer to \cite{ramsaysilvermanbook}, for further details. \\
Let $(\Omega, \mathfrak{F}, \mathbb{P})$ be a probability space, and, for $\omega\in \Omega$, denote by $X(\omega)$ an element of the functional Hilbert space $L^2(I)$, where $I$ is a closed interval in $\mathbbm{R}$. Suppose we are provided with a functional dataset $\{x_i\}_{i=1}^{n}$, independent realizations of $X$, and we aim at reducing its dimensionality. Formally, the first functional principal component is the function $\beta^*(t)\in L^2(I)$ that maximizes the \textit{penalized sample variance}:
\begin{equation}
    \beta^*(t)=\operatorname*{argmax}_{\beta}\frac{\widehat{\operatorname*{Var}}(\langle\beta,X\rangle_{L^2(I)})}{||\beta||_{L^2(I)}+\lambda Pen(\beta)}\text{ ,}
\end{equation}

\noindent
where $\widehat{Var}(\cdot)$ is the sample variance and, typically, $Pen(\beta)=||\beta''||_{L^2(I)}^2$ and $\lambda>0$ is the penalization parameter (\cite{silverman96}). We can compute the subsequent principal components, with the additional constraint of being orthogonal to previously found components, by solving the equivalent eigenvalue problem (\cite{silverman96}).  FPCA allows to gain valuable insights into the inherent structures and dynamics of the underlying processes. Nevertheless, when the objective is to utilize a functional dataset for predicting a new object, employing fPCA does not guarantee that the extracted components are the most appropriate for forecasting purposes, e.g. when the variance explained by any covariate is masked by a high global variance.

\subsection{Functional Fisher's Discriminant Analysis}
\label{foFD}
When the aim is supervised classification, a well-known dimensionality reduction technique is Fisher's Discriminant Analysis, by \cite{fisher36}. As PCA, this technique was extended to the functional framework in \cite{ramsaysilvermanbook}, based on \cite{hastie95}, in which Penalized Discriminant Analysis is discussed and the equivalence with Discriminant CCA and Optimal Scoring is demonstrated for categorical labels. Let $X$ be a random function as in the previous Section (\ref{fPCA}). Let us suppose that our dataset consists of pairs $\{(x_i(t),c_i)\}_{i=1}^{n}$, where, for any $i$, $x_i$ is a realization of $X$ and $c_i\in\{c_0,c_1,c_2,...,c_{n_\mathcal{C}}\}$ is the realization of a categorical variable. The first functional Fisher discriminant component is a function $\beta^*\in L^2(I)$ which maximizes the between-class sample variance, ensuring that the projected classes are well-separated, while simultaneously minimizing the within-class sample variance to maintain compactness within each class, formally:
\begin{equation}
   \beta^*(t)=\operatorname*{argmax}_{\beta}\frac{{\widehat{\operatorname*{Var}}_B(\langle\beta,X\rangle_{L^2(I)})}}{{\widehat{\operatorname*{Var}}_W(\langle\beta,X\rangle_{L^2(I)})}+\lambda Pen(\beta)}\text{ ,}
    \label{eq_FDA}  
\end{equation}

\noindent
where $\operatorname*{Var}_B$ is the between-class sample variance, and $\operatorname*{Var}_W$ is the within-class sample variance, i.e.:
\begin{equation}
    \begin{cases}
            \widehat{\operatorname*{Var}}_B(\langle\beta,X\rangle_{L^2(I)})=\displaystyle\frac{1}{n_\mathcal{C}}\sum_{c\in\{c_0,...,c_{n_\mathcal{C}}\}}{n^{(c)}(\langle\beta,\bar{x_c}-\bar{x}\rangle)^2}\\
    \widehat{\operatorname*{Var}}_W(\langle\beta,X\rangle_{L^2(I)})=\displaystyle\frac{1}{n}\sum_{i\in\{1,...,n\}}{(\langle\beta,x_i-\bar{x}_{c_i}\rangle)^2}\\
    \bar{x}_c=\frac{1}{n^{(c)}}\displaystyle\sum_{i:c_i=c}x_i \text{\hspace{1cm}and\hspace{1cm}} \bar{x}=\displaystyle\frac{1}{n}\sum_{i\in\{1,...,n\}}{x_i}\\
    n^{(c)}=|\{i:c_i=c\}| \text{\hspace{1cm} for any \hspace{1cm}} c\in\{c_0,c_1,..,c_{n_\mathcal{C}}\}
    \end{cases}
\end{equation}
\noindent
This technique allows to extract the component that best discriminates among the classes. The Fisher's concept can be modified to be used with an ordinal label, rather than categorical, extending the idea of (\cite{mcfee2014_segments}) to the functional framework, i.e. substituting the classical sample between covariance $\operatorname*{Var}_B$ with:
\begin{equation}
\widehat{\operatorname*{Var}}^{O}_{B}(\langle\beta,X\rangle)=\sum_{c=0}^{n_{\mathcal{C}}-1}n^{(c)}(\langle\beta,\bar{x}_c-\bar{x}_c^+\rangle)^2+n^{(c+1)}(\langle\beta,\bar{x}_{c+1}-\bar{x}_c^+\rangle)^2   
\end{equation}
\noindent where $\bar{x}_c^+$ is the mean vector of the $x_i$'s belonging to the union of the two consecutive levels $c$ and $c+1$. With this modification, the discrimination is maximized between consecutive levels instead of among all the possible classes. In the following, we will refer to this method with the name \textit{functional-ordinal Fisher Discriminant} (foFD). When the underlying data distribution is approximately Gaussian and the class covariances are assumed to be equal, foFD provides an optimal linear transformation for discrimination. In view of this, foFD could be considered as a competitor method of that proposed in this work. However, we do not pursue this approach further as it is subject to limitations, e.g.,  when data do not come from a Gaussian process (\cite{johnsonwichernbook}), when the variance within each class is maximized along the same direction as the variance between classes (since the same direction maximizes the numerator and the denominator of the fraction in Equation \ref{eq_FDA}) or when the groups are separated by very different distances. The latter issue holds also net of the modification of the between variance definition for ordinal labels, proposed by \cite{mcfee2014_segments}. 

\section{Functional-ordinal Canonical Correlation Analysis}
\label{foCCA_met}
We here present the functional-ordinal Canonical Correlation Analysis, which aims to provide a dimensionality reduction of a functional dataset, as to maximize the discrimination ability, in the reduced space, among the levels of an ordinal variable. 
\subsection{The maximization problem}
Suppose that we are provided with $n$ independent realizations of the functional variable $X$ and the vector $Y$, which is defined in Equation \ref{defY} starting from an ordinal random variable $\mathcal{C}$. Thus, our samples consists of pairs $\{x_i(t),y_i\}_{i=1}^{n}\in L^2(I)\times\{0,1\}^{n_{\mathcal{C}}}$. The main objective is to be able to predict the $Y$, i.e. the level $\mathcal{C}$, knowing the functional variable $X$. We can reduce the dimensionality of the functional part of the dataset while extracting the features that best discriminate among different levels, adapting Canonical Correlation Analysis to the functional-ordinal context. More precisely, foCCA looks for $\beta^*\in L^2(I)$ and $\theta^*\in \mathbbm{R}^{n_{\mathcal{C}}}$ that maximize the sample correlation between the following random variables

\[\langle\beta,X\rangle_{L^2(I)} \text{\hspace{1cm}and\hspace{1cm}} \langle\theta, Y\rangle_{\mathbbm{R}^{n_\mathcal{C}}}\].

\noindent
We call $\beta(t)$ the functional canonical component and $\theta$ the ordinal canonical component, while $\langle\beta,X\rangle_{L^2(I)}$ and $\langle\theta,Y\rangle_{\mathbbm{R}^{n_\mathcal{C}}}$ are, respectively, the functional and the ordinal scores. Formally, in analogy to \cite{leurgans93}, where both datasets are functional, we want to find the functional object $\beta^*(t)$ and the vector $\theta^*$ that maximize the \textit{penalized squared sample correlation}:\\
\begin{equation}
    ccorsq_{\lambda_1,\lambda_2}(\beta,\theta)=\frac{\widehat{\operatorname*{Cov}}(\langle\beta,X\rangle_{L^2(I)},\langle\theta,Y\rangle_{\mathbbm{R}^{n_\mathcal{C}}})^2}{\biggl(\widehat{\operatorname*{Var}}(\langle\beta,X\rangle_{L^2(I)})+\lambda_1 Pen(\beta) \biggr)\biggl(\widehat{\operatorname*{Var}}(\langle\theta,Y\rangle_{\mathbbm{R}^{n_\mathcal{C}}}) +\lambda_2 Pen(\theta)\biggr)}
    \label{opt}
\end{equation}
\noindent
where $\widehat{Cov}(\cdot)$ and $\widehat{Var}(\cdot)$ are, respectively, the sample covariance and variance, $Pen(\beta)$ is a smoothing penalization and $Pen(\theta)$ is an elastic net penalization, with the respective penalty parameters $\lambda_1>0$, $\lambda_2>0$. The need for penalization is deeply discussed in \cite{silverman96}. In our work, for the functional canonical component the penalization is $Pen(\beta)=||\beta''||_{L^2(I)}^2$, typically used in smoothing spline regression (\cite{ramsaysilvermanbook}), and for the ordinal canonical component $Pen(\theta)=||\theta||^2$, i.e. the Ridge penalization (\cite{hoerl70}). Given that the described optimization problem is equivalent to a generalized eigenvalue problem, we can derive the first pair of canonical components $(\beta^{(1)}(t),\theta^{(1)})$ corresponding to the largest eigenvalue $\rho^{(1)}$, the second pair $(\beta^{(2)}(t),\theta^{(2)})$ corresponding to the second largest eigenvalue $\rho^{(2)}$, and so forth, up to a desired number of components $m$.  In the next Section \ref{discrete} we elucidate the computational tools employed to derive the pairs of canonical components, providing a detailed resource. 

\subsection{Discretization of the maximization problem}
\label{discrete}
We here enter into the details of the discretizaton of the dataset and of the optimization problem. Moreover, we provide the subsequent derivation of the generalized eigenvalue problem. These steps allows the computation of the sequence of $m$ canonical correlation components. In the following, we assume both the functional and the ordinal data to be empirically centered to facilitate the computations. Net of the centering, the maximization problem (\ref{opt}) is equivalent to the following:
\small{\begin{equation}
    (\beta^*,\theta^*)=\operatorname*{argmax}_{\beta,\theta} \frac{\displaystyle\frac{1}{n}\sum_{i=1}^{n}\langle\beta,X_i\rangle_{L^2(I)}\langle\theta,y_i\rangle_{\mathbbm{R}^{n_\mathcal{C}}}}{\displaystyle\bigg(\frac{1}{n}\sum_{i=1}^{n}\langle\beta,X_i\rangle_{L^2(I)}^2+\lambda_1 \langle\beta,\beta\rangle_{L^2(I)}^2\bigg)\bigg(\frac{1}{n}\sum_{i=1}^{n}\langle\theta,y_i\rangle_{\mathbbm{R}^{n_\mathcal{C}}}^2+\lambda_2\langle\theta,\theta\rangle_{\mathbbm{R}^{n_\mathcal{C}}}^2\bigg)}
    \label{opt2}
\end{equation}}
Let us focus on the discretization of the functional dataset, as the ordinal dataset is already finite dimensional. Suppose that we can represent, possibly after a smoothing regression, each functional datum $x_i(t)$, where $1\leq i \leq n$, through a finite basis expansion, i.e.:
        \[
        x_i(t)=\sum_{j=1}^{n_J}a_{ij}\phi_j(t)
        \]
        where $a_{ij}$ is the coefficient associated with the basis function $\phi_j(t)\in L^2(I)$. 
If we define the following matrices:
\begin{itemize}
    \item  $\mathcal{A}=\{a_{ij}\}_{ij}$, where $a_{ij}$ is the coefficient of the i-th unit associated to the j-th spline, in centered dataset $\{x_i(t)\}_i$.
    \item $\mathcal{D}=\{d_{il}\}_{il}$ where $d_{il}$ is the coefficient of the i-th unit associated to the c-th basis vector, in the centered $\{y_i\}_i$.
    \item $V_{12}=n^{-1}{\mathcal{A}^T\mathcal{D}}$, $V_{11}=n^{-1}{\mathcal{A}^T\mathcal{A}}$, $V_{22}=n^{-1}{\mathcal{D}^T\mathcal{D}}$. 
    \item $K=\biggl\{\displaystyle\int_{I}{\frac{\partial^2\phi_j}{dt^2}\frac{\partial^2\phi_k}{dt^2}}\biggr\}_{j,k}$
    \item $J=\biggl\{\displaystyle\int_{I}{\phi_j\phi_k}\biggr\}_{j,k}$
\end{itemize}
By setting $\beta(t)=\sum_{j}b_j\phi_j(t)$, the maximization problem in Equation \ref{opt2} can be discretized as follows:
\begin{equation}
    (b^*,\theta^*)=\operatorname*{argmax}_{b,\theta} \frac{(b^T JV_{12} \theta)^2}{(b^T JV_{11}J b + \lambda_1 b^TKb)(\theta^T V_{22} \theta + \lambda_2 \theta^T\theta)}
\end{equation}
which is in turn equivalent to maximize the value $(b^T JV_{12} \theta)$ imposing $(b^T JV_{11}J b + \lambda_1 b^TKb)=(\theta^T V_{22} \theta + \lambda_2 \theta^T\theta)=1$. Hence, the optimization problem in (\ref{opt}) is equivalent to the following generalized eigenvalue problem (\cite{ramsaysilvermanbook}): 
\begin{equation}
    \begin{bmatrix}
0 & JV_{12} \\
V_{21}J & 0 
\end{bmatrix}
\begin{bmatrix}
b\\
\theta
\end{bmatrix}
=\rho
\begin{bmatrix}
JV_{11}J+\lambda_1 K_1 & 0 \\
0 & V_{22}+\lambda_2 I
\end{bmatrix}
\begin{bmatrix}
b\\
\theta
\end{bmatrix}
\end{equation}

\noindent
The eigenvector $[{b^*}^T,{\theta^*}^T]$ with highest eigenvalue $\rho$ allows to reconstruct the first canonical component as $\beta^*(t)=\sum_j{b_j^*}{\phi_j(t)}$, while the second block $\theta^*$ directly provides the first ordinal canonical component. 

\subsection{Interpretation of canonical components}
\label{theta}
In this section, we enter into the interpretation of the canonical components derived from the maximization problem introduced above. By elucidating the interpretation of these components, we seek to unravel valuable insights into the relationships between the functional and the ordinal datasets. Recall that foCCA computes the function $\beta^*\in L^2(I)$ and the vector $\theta^*\in\mathbbm{R}^{n_{\mathcal{C}}}$ that maximize the correlation between the functional scores $\langle\beta,X\rangle_{L^2(I)}=\int_{I}{\beta(t)X(t)dt}$ and the ordinal scores $\langle\theta,Y\rangle=\sum_{j=1,...,n_\mathcal{C}}{\theta_j Y_j}$. The functional canonical component $\beta^*(t)$ is the weight given to the functional datum $X(t)$ across time, in the functional score computation. This function makes clear which are temporal intervals that are valuable for discrimination and highlights how they influence the functional score that optimizes the prediction of the ordinal variable. For what concerns the ordinal canonical component, notice that the suitable representation of $\mathcal{C}$ through $Y$ allows to make flexible the distance between consecutive levels in the correlation analysis, exploiting the ordinal nature of the variable $\mathcal{C}$. Indeed, $\theta^*$ contains the components representing the step between two consecutive ordinal variables. For instance, if we consider two realizations of $Y$, such that $c_1=\hat{c}-1$ and $c_2=\hat{c}$ ($\hat{c}\in\{1, ..., n_C\}$), i.e. \[y_1=\sum_{c=1}^{\hat{c}-1}{e_c} \text{\hspace{1cm}and\hspace{1cm}} y_2=\sum_{c=1}^{\hat{c}}{e_c}\text{ ,} \] then the difference between the two ordinal scores is $({\theta^*}^T y_2 - {\theta^*}^T y_1 ) = {\theta^*}^T (y_1-y_2)={\theta^*}^T e_{\hat{c}}=\theta_{\hat{c}}$. It is important to note that even after centering the dataset $Y$, the disparity between two ordinal observations remains constant. Consequently, the interpretation of the coefficients in $\theta^*$ remains unaffected by this centering process.

\subsection{A note on penalty parameter selection}
\label{penalty}
Since our final goal is to optimize the dimensionality reduction in terms of classification power, we select the parameters with a cross-validation method: we fix the K-fold Cross Validated Mean Absolute Error ($MAE_{CV}$) as the loss function, i.e.:
\begin{equation}
    MAE_{CV}=\sum_{k=1}^{K}\sum_{i\in F_k} \frac{|c_i-\hat{c}^{(-F_k)}_i|}{n_{F_k}} \text{ ,}
\end{equation}
where $\hat{c}^{(-F_k)}_i$ is the level associated with the nearest centroid (calculated by excluding $F_k$) within the space of the first $m$ components, where $m$ is fixed. Here, we identify the penalty parameters corresponding to the minimum of the smoothed loss curve (or surface). As revealed by sensitivity analyses, smoothing is essential as the narrow selection of the penalty parameter based solely on the raw loss curve (or surface) may be influenced by the subdivision into K-folds. To elaborate further, the loss is smoothed through smoothing cubic splines regression (with smoothing parameter selected via GCV, \cite{wood04}); since in foCCA we have to select two parameters, we consider smoothing regression with tensor product of cubic splines (\cite{wood15}). 

\section{Simulation Study}

In this section, we present a simulation study conducted to assess the performance of the dimensionality reduction method foCCA, and comparing it with the performance of fPCA and foFD, under specific conditions, via Monte Carlo simulations. The foCCA method, by design, is expected to outperform competitors in identifying functional components that effectively differentiate between various levels of an ordinal variable, even in cases where the intervals between consecutive levels are not uniformly distanced. Consequently, we conduct comparative analyses across two distinct simulated scenarios: (a) A scenario where the variance attributed to the ordinal variable is lower than the overall dataset variance, which, in particular, poses challenges for fPCA; (b) A scenario where the successive levels of the ordinal variable exhibit non-uniformity in distance, which, in principle, poses challenges to all the methods.
\vspace{0.5cm}\\
This structured approach allows us to systematically evaluate and compare how each method responds to challenges specific to the underlying data structures, providing information into their respective strengths and limitations in practical applications.

\label{sim_study}
\subsection{Simulated scenarios}
\label{simulations}
\begin{figure}[H]
    \centering    \includegraphics[scale=0.5]{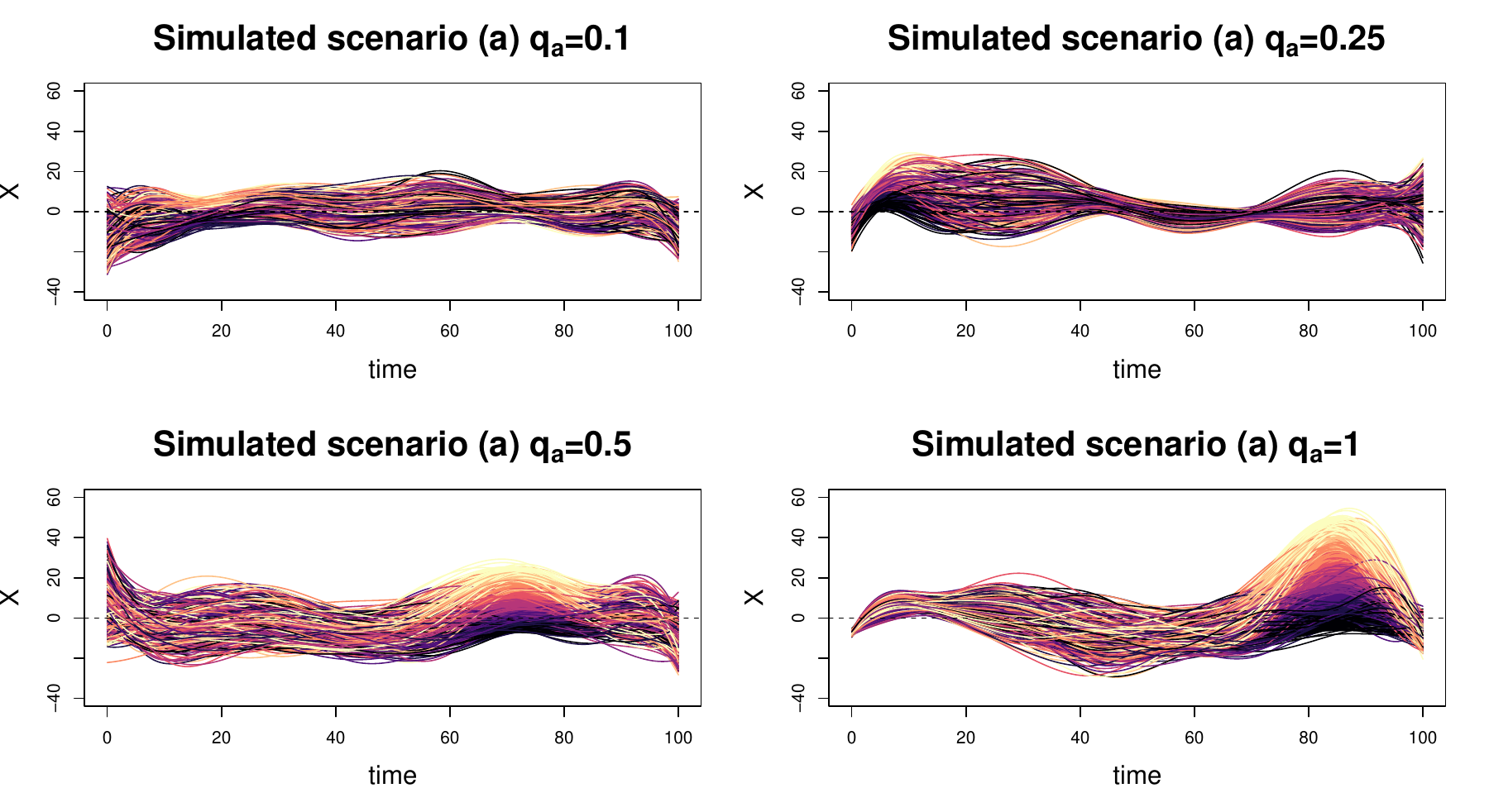}
    \caption{Examples of simulated scenarios of type (a), depending on $q_a$}
    \label{scena}
\end{figure}

The simulated functional dataset consists of $n=1000$ functional data built on a basis of $J=10$ cubic B-splines, with equispaced knots and range $I=[0,100]$, where the coefficients of the splines are independently randomly sampled from a distribution that depends on the scenario we aim at creating. The number of levels is fixed equal to $(n_\mathcal{C}+1)=9$, as in the case study in Section \ref{motivating}, and for any unit $i\in\{1,..,n\}$, the level $c_i$ is randomly sampled from the discrete uniform distribution over $\{0,1,2,...,8\}$. 
\vspace{0.5cm} \\
In the scenario (a), a singular Monte Carlo simulation consists of a dataset of realizations $(x_i(t),c_i)_{i=1}^{n}$. For any unit $i\in\{1,...,n\}$, $x_i(t)=\sum_{j=1}^{J}{a_{ij}\phi_j(t)}$, where $a_{ij}$ is the coefficient of the $i-th$ unit associated with the $j-th$ spline of the basis. The coefficients $\{a_{ij}\}_{i=1}^{n}$ are independently sampled from the normal distribution $\mathcal{N}(\mu_j+h_{ij},\sigma_j)$, where, for any spline $j$, $\mu_j$ is the general mean, $h_{ij}$ is a level-specific increment and $\sigma_j$ is the standard deviation. For any Monte Carlo simulation, $\{\mu_j\}_{j=1}^{J}$ are i.i.d sampled from $U([-10,10])$, and $\{\sigma_j\}_{j=1}^{J}$ are i.i.d sampled from $U([0,10])$. The level-specific increment $h_{ij}$ is equal to $\sum_{k=0}^{c_i}\gamma(k)$ if $j=s$, and it is equal 0 otherwise. For any Monte Carlo simulation the spline $s=\displaystyle\operatorname*{argmin}_{j\in\{1,...,J\}}{\{\sigma_j\}}$ is the spline along which we generate the ordinal pattern, defined by the increment $\gamma(k)\sim\mathcal{N}(q_a\cdot 10,1)$ for any $k\in\{1,...,n_\mathcal{C}\}\text{ and } \gamma(0)=0$. In particular, $q_a$ denotes a parameter that controls the severity of the perturbations across levels applied in the generated scenario. The reason why we generate different $\{\mu_j\}_{j=1}^{J}$, $\{\sigma_j\}_{j=1}^{J}$ and $\gamma$ for each Monte Carlo simulation, is that we want to be sure that all the methods are challenged: indeed, for any simulated dataset we change the mean function, the direction along which the ordinal pattern is built, the direction along which the global variance is the highest and the mutual distances between the consecutive levels. Hence, the methods are compared on an exhaustive and wide scenario. Notice that, when $q_a=0$ the increment is null for each spline $j$, i.e. the variance explained by the ordinal variable is zero along the entire domain. On the contrary, when $q_a>0$, along the spline $s$ the coefficients have a mean increment $(q_a\cdot 10)c_i$, i.e. the larger $q_a$, the higher the expected ratio between the variance explained by the ordinal variable and the global variance. See Figure \ref{scena} to see some examples of scenarios of type (a).
\vspace{0.5cm} \\
\noindent In the scenario (b) the successive levels of the ordinal variable exhibit a non-uniform separation, both in terms of distance and in terms of component along which the distance is maximized. Each Monte Carlo simulation consists of a dataset of realizations $(x_i(t),c_i)_{i=1}^{n}$, with $x_i(t)=\sum_{j=1}^{J}{a_{ij}\phi_j(t)}$, where $a_{ij}$ are independently sampled from $\mathcal{N}(\mu_{ij},1)$. In particular, $\mu_{ij} = \mu_j^{(0)} + q_b\mathbbm{1}_{\{ c_i \geq 1\}}(\sum_{1\leq c\leq min(4,c_i)} \mu_j^{(1)}) + \mathbbm{1}_{\{ c_i \geq 5\}} \sum_{5\leq c\leq c_i } \mu_j^{(2)}$, where for any Monte Carlo simulation $\mu_j^{(k)}$ are i.i.d sampled from $\mathcal{U}(-1,1))$. 
Hence, the mean $\mu_{ij}$ is characterized by a global mean term $\mu_j^{(0)}$ and by two additional terms that depend on the level  $c_i$ associated with the unit. More precisely, $\mu_j^{(1)}$ is the increment applied to $\mu_{ij}$ for each level $k\leq c_i$ with $k = 1,...,4$, while $\mu_j^{(2)}$ is the analogous increment among contiguous levels but for the highest levels $k\geq c_i$ with $k = 5,...,8$. In practice, let us consider to start with $\mu_{ij}=\mu_{j}^{(0)}$. The value $q_b$ is the ratio between the increments of the highest levels and the one of the lowest levels, hence it represents the severity of the Scenario (b). A lower value of $q_b$ indicates less uniform separation between consecutive levels, illustrating greater disparity in the increments for higher levels compared to lower levels. 

\begin{figure}[H]
    \centering    \includegraphics[scale=0.5]{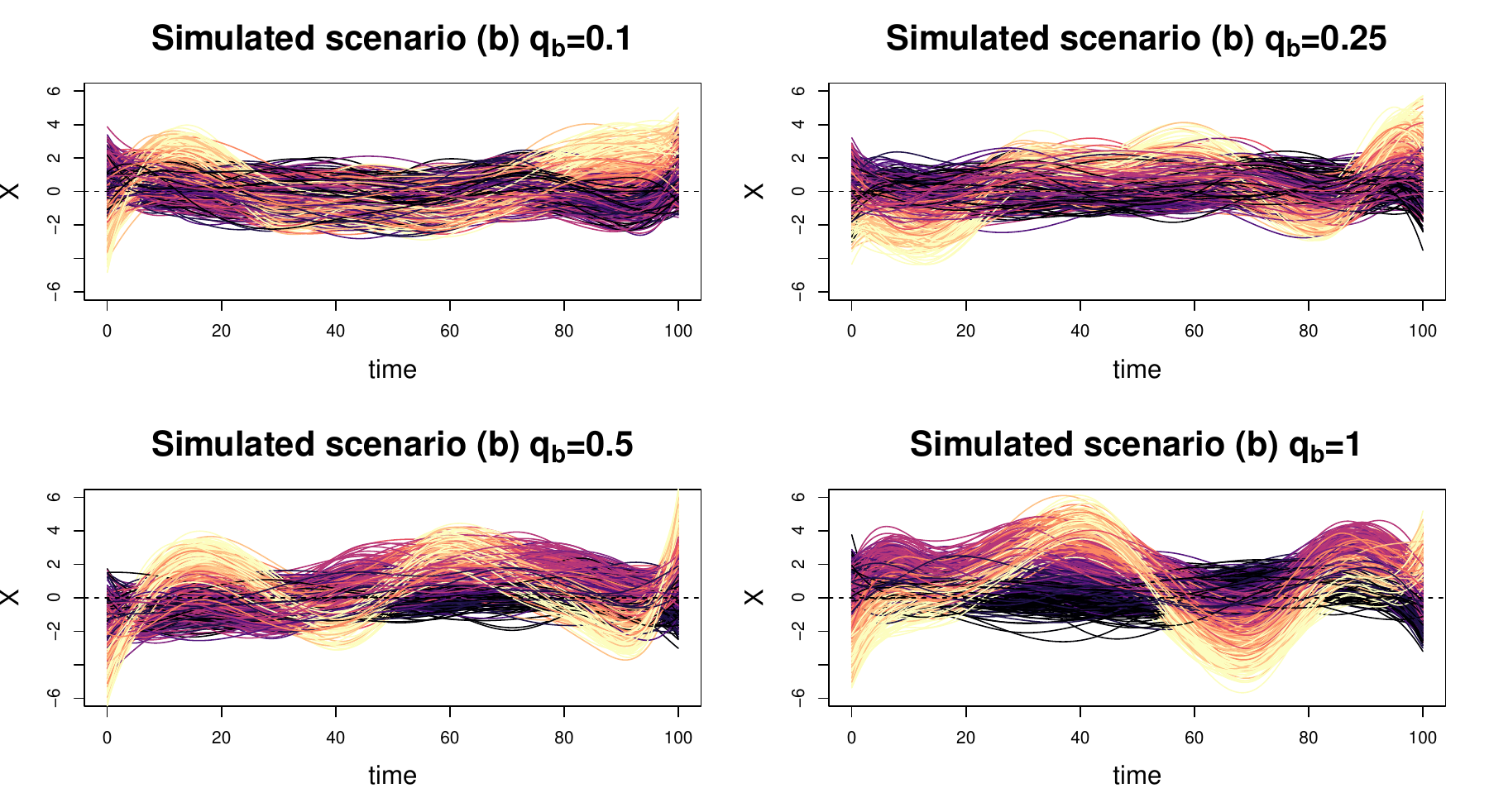}
    \caption{Examples of simulated scenarios of type (b), depending on $q_b$}
    \label{scenb}
\end{figure}

\subsection{Results}
\begin{figure}[H]
\centering\includegraphics[scale=0.6]{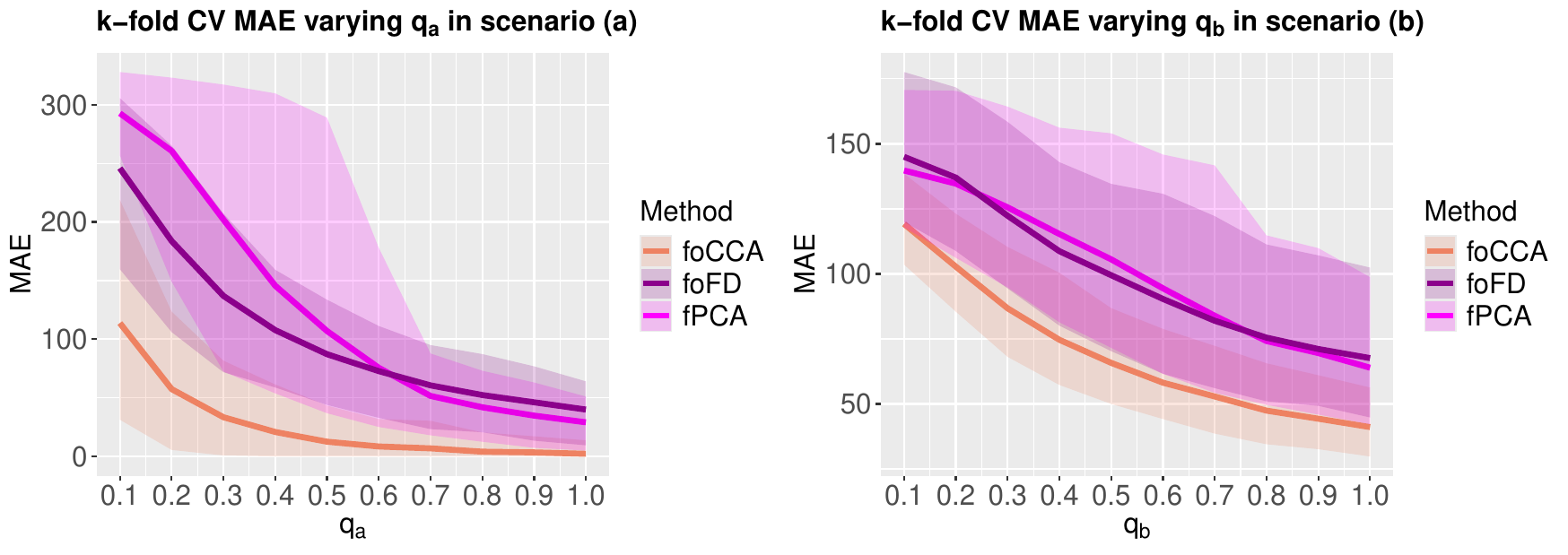}
    \caption{Monte Carlo simulations: the line represents the mean values, while the bands' boundaries are the pointwise quantiles $5\%$ and $95\%$}
    \label{MC_sim}
\end{figure}

In this Section, we aim to estimate the performance of foCCA via Monte Carlo simulations, built by varying the ratio $q_a$, i.e. generating scenarios of type (a), and $q_b$, i.e. generating scenarios of type (b). We compare the results with the ones obtained through fPCA and foFD. We run $M=500$ simulations for each scenario, letting $q_a$ and $q_b$ vary in the range $\{0, 0.1, 0.2,...,1\}$., and we evaluate the K-fold MAE, defined in Section \ref{penalty}, with $K=5$. By doing this, we can see how the severity of the scenario affects the models. \\
Results are depicted in Figure \ref{MC_sim}. In the scenario (a) we can observe that the performances of all the three methods improve as $q_a$ increases. This is due to the fact that, the higher $q_a$, the clearer the separation of the levels (see also Figure \ref{scena}). Indeed, a high ordinal variability, with respect to the global variability, drive all the methods towards the identification of components that intrisically separate the levels. However, foCCA proves consistently better than the competitors across all the severity levels. Notice that, when $q_a$ is high, foFD is the worst method. A possible reason for this result relies in the fact that the direction that maximizes the ordinal variance is, by construction, the same as the one that maximizes the global variance. Contrarily, when $q_a$ is small, the performance of fPCA is the poorest. This phenomenon can be explained by the fact that, in this scenario, the ordinal variance is negligible with respect to the global variance, hindering the identification of discriminant directions through fPCA. \\
Let us now focus on scenario (b). In this scenario, whatever the value of $q_b$, the levels are not equally separated. However, as $q_b$ grows, the recognition of the lowest levels should, in principle, be possible for each method. We can observe from the Monte Carlo simulations, in Figure \ref{MC_sim}, that as $q_b$ increases, the error of all the methods decreases. However, for each $q_b\in[0,1]$, while fPCA and foFD tends to overlap in terms of MAE, the difference between these two methods and foCCA is always high in mean. In conclusion, foCCA is able to keep a significantly lower error than competitors in both scenarios proposed. The methods behave in a similar way only when there is an ordinal variance comparable to the global one, with levels equispaced along the same components.

\section{A case study: analyzing biological interactions from biosensor video signals}
\label{motivating}

\subsection{Optical biosensors data: background}
In recent decades, optical biosensors have become increasingly important in a wide variety of areas such as biomedical diagnostics (\cite{sharma21}), drug discovery (\cite{cooper06}), food safety (\cite{narsaiah12}), or environmental monitoring (\cite{sharma23,long13}). To enhance the time-efficiency and sustainability of optical biosensors, imaging techniques have been developed to measure multiple interactions between the receptor (such as an enzyme, nucleic acid, an antibody,
tissue, organelles or even whole cell) and the analyte (e.g. an antigen of interest) in parallel. Due to these challenges and progressions in complexity and design (\cite{narlawar24,gauglitz10}), the generated datasets are becoming high-dimensional, yielding  time-consuming and memory-intensive data processing. 

In this study, a reflectometric imaging sensor  is used to track the progression of the interaction between an antibody and an antigen immobilized onto a transducer, based on a video of the light intensity.  The primary statistical challenge lies in analyzing temporal profiles of light intensity within Regions of Interest (ROIs), where the antibody is present with a nominal concentration level. In particular, our interest lies in analyzing the light intensity evolution depending on the level of antigen concentration, and therefore, in predicting the level of antigen concentration, based the optical signal, in a new ROI.

\noindent 
\subsection{Preprocessing}
\label{preproc}
\begin{figure}
    \centering
    \includegraphics[scale=0.34]{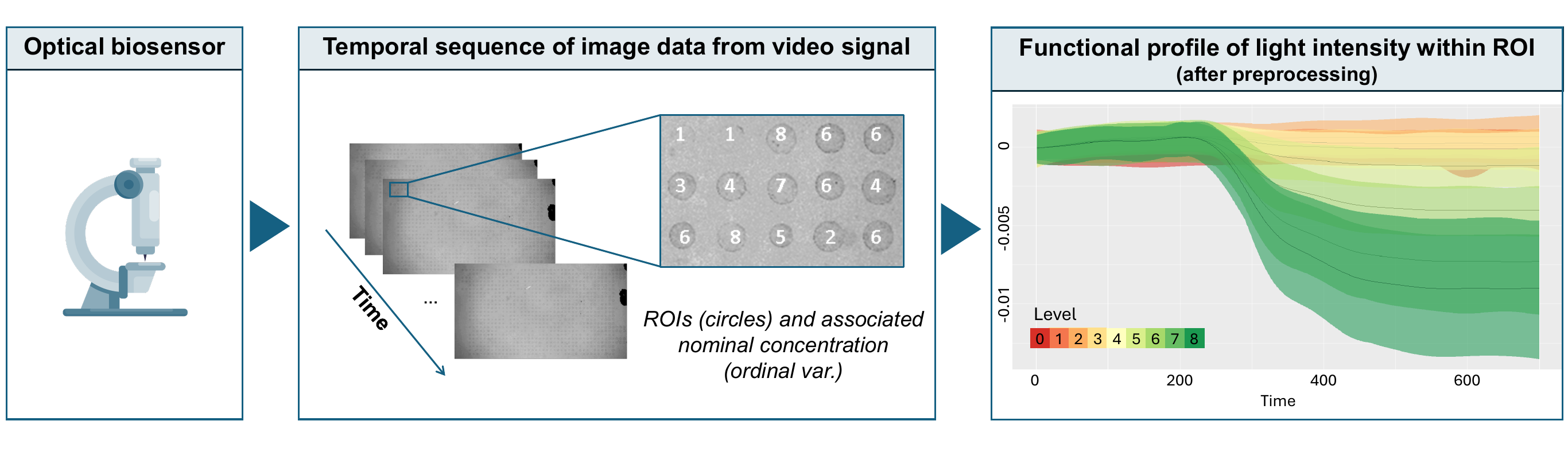}
    \caption{Biosensor data, a zoom on a subset of ROIs labeled with a nominal concentration level. On the right, the functional boxplot of the smoothed log-ratio, stratified by concentration level. Coloring of the boxplots is consistent with that of the associated levels (on the right). Note that levels 0 denotes the control group, only levels $c_i>1$ display a spot on the image, and only levels $c_i>2$ presents a positive antigen concentration.}
    \label{intro_image}
\end{figure}

The available dataset consists of a video signal, recording one image per second, captured through a reflectometric sensor. Each frame represents the light intensity of the reflected light, measured at a time instants. Here, 1035 spots--each containing a certain concentration of the antigen t-BSA--are arranged in a regular grid on the sensor surface, identifying 1035 regions of interest (ROIs). In practice, some spots may not exist or contain minimal reagent concentration. Note that visually detecting the presence of a spot is straightforward, as it appears darker than the background in the reflectometric image. 
Figure \ref{intro_image} reports a schematic representation of the dataset.

Preprocessing of the video signal consists of light cleaning and subsequent construction of the ordinal-functional dataset, where the statistical unit is the ROI and the target variables are (i) a functional data point that represents the temporal evolution of the light intensity and (ii) an ordinal variable that represents the level of concentration of reagent.

For each ROI, the nominal concentration level is associated with an ordinal label. In order to detect magnitude outliers among the ROIs, we eliminate exogenous sources of light from the reflectometric images, such as external lights and irregularities in the surface coating. In our case study, it is realistic to assume that the exogenous sources of light are constant over time. We model the logarithm of light intensity at time 0 as a Generalized Additive Model (GAM, \cite{hastietibshirani}), using as predictor the pixels's spatial coordinates $(x,y)$, with a nonlinear contribution, and the categorical information of whether or not the pixel belongs to a ROI. By dividing the light intensity by the model-based expected light disturbance (or by subtracting in the case of their logarithms) on the reflectometric images, we effectively minimize variations in light that do not stem from the presence of the antigen spot. Net of the light cleaning, we can easily check if a pixel belongs to random spots or determine if a randomly sampled pixel is located within or outside an ROI. 
\begin{figure}[H] 
\centering\includegraphics[scale=0.35]{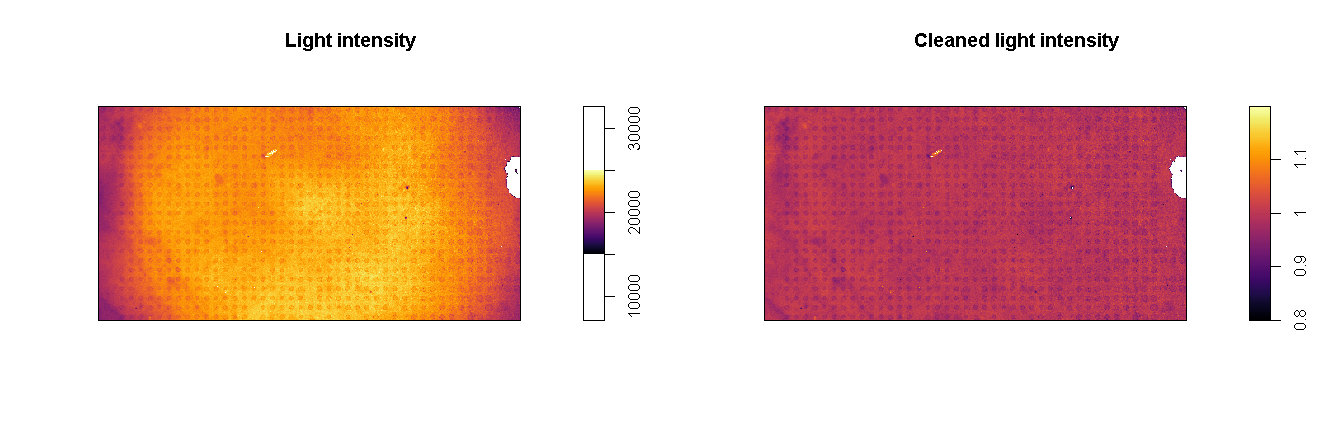}
    \caption{Light intensity vs cleaned light intensity at time 0}
    \label{adj_light}
\end{figure}
\vspace{0.5cm}
\noindent
The light intensity, ranging from $0$ to $2^{16}-1$, exhibits a compositional nature. To conduct standard statistical analyses like dimensionality reduction, modeling, or clustering, it is crucial to work with absolute values, rather than the relative information associated with the compositional nature of the data (\cite{composit15}). Various transformations for compositional data have been proposed in the literature to ensure meaningful and interpretable results. Among these, the log-ratio approach, initially introduced by \cite{aitchison}, stands out as one of the most popular methodologies. Following these principles, we construct a functional datum representing the light intensity of the $i$-th ROI as follows. We subtract the GAM model from the logarithm of light intensity, obtaining a cleaned, time-varying log-light intensity, $f_j(t)$, for each pixel $j$ in the image $D$. For each ROI $i$, we compute the $5\%$-trimmed mean of the cleaned signals $f_j(t)$ corresponding to $N=100$ randomly extracted pixels within the ROI, resulting in $\bar{f}_i^{ROI}(t)$. The trimming is based on the Modified Band depth, introduced by \cite{lopez09}. This ensures robustness in estimating the mean function within the ROI, which is then taken as representative of the signal for that ROI. Additionally, we sample $N=100$ pixels in the neighborhood of each ROI $i$ and compute the $5\%$-trimmed mean $\bar{f}_i^{BG}(t)$ to robustly estimate the mean in the background area adjacent to ROI $i$. After conducting a sensitivity analysis regarding the quantity of points sampled within and outside the region of interest, it became apparent that selecting more than 100 points does not result in a noticeable enhancement. To account for the compositional nature of the signals, we compute the log-ratio of the light intensity: 
\begin{equation}
   {x}_i(t)=(\bar{f}_i^{ROI}(t)-\bar{f}_i^{ROI}(0))-(\bar{f}_i^{BG}(t)-\bar{f}_i^{BG}(0)) \text{ .}
   \label{logratio}
\end{equation}
We then smooth the log-ratio (\ref{logratio}) using smoothing splines regression (\cite{splines}), employing 20 cubic splines with equally distributed knots and a penalty parameter of $10^4$ selected via GCV to obtain a meaningful functional profile. In conclusion, each ROI $i$ is associated with a functional datum $x_i(t)$ and a concentration level $c_i$ of the reagent. Figure \ref{intro_image} displays (on the right) the functional boxplot of the log-ratio of the light intensity, colored according to the associated level of antigen concentration. The final dataset consists of the logratio profiles dataset deprived of the level-specific functional outliers, recognised through the Modified Band Depth (\cite{lopez09}). \\
\subsection{Dimensionality reduction}
\label{case_study}
\begin{figure}[H]
    \centering
    \includegraphics[scale=0.6]{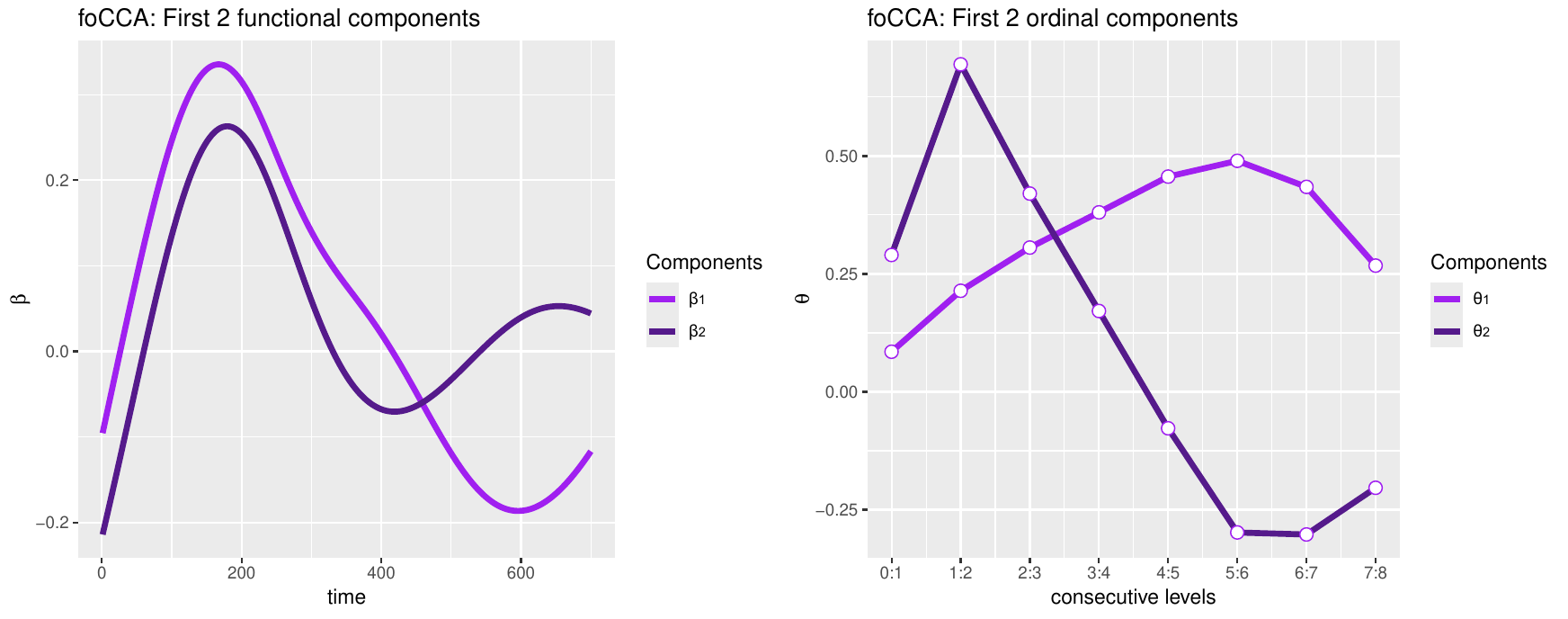}
    \caption{The first two components of foCCA}
    \label{focca_comp}
\end{figure}

\begin{figure}[H]
    \centering
    \includegraphics[scale=0.8]{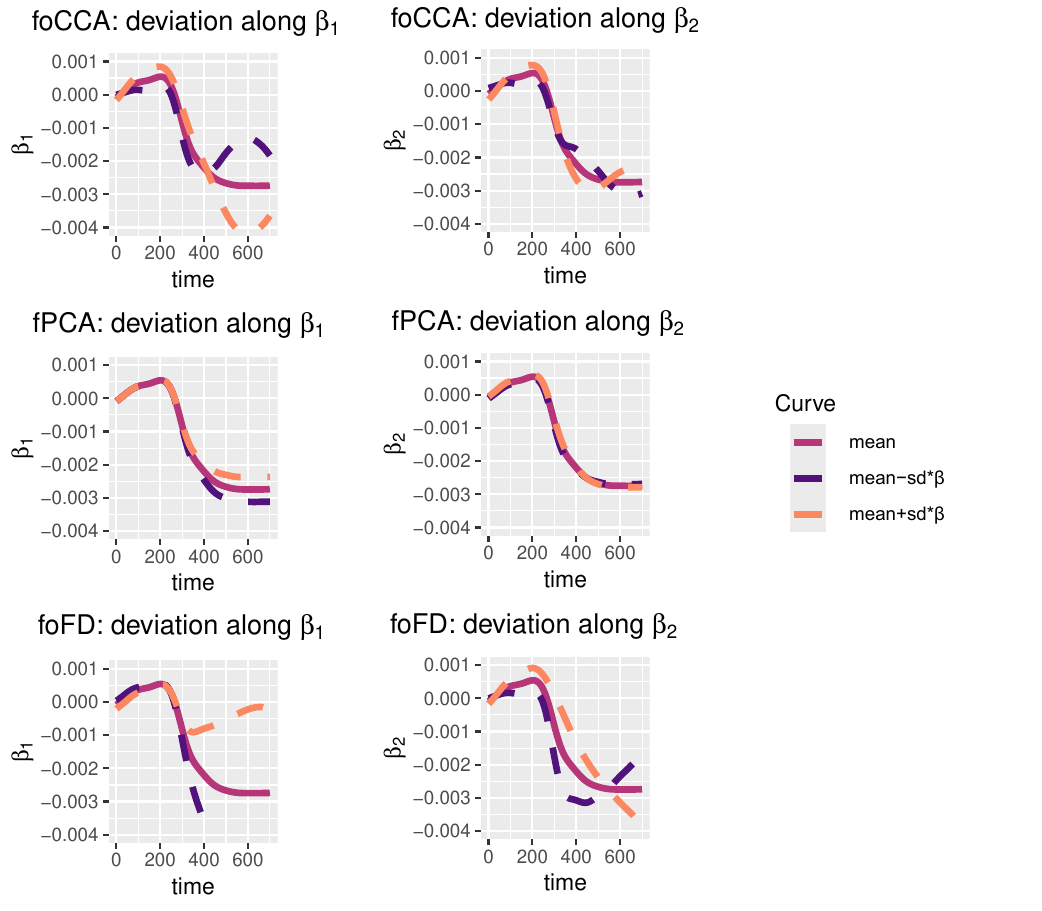}
    \caption{Deviations along the first two functional components}
    \label{deviations}
\end{figure}

\noindent
In this section, we show the results obtained by applying the method foCCA, presented in Section \ref{foCCA_met}, to the preprocessed dataset introduced in Section \ref{preproc}. From a biophysical standpoint, levels 0 and 1 do not have intrinsic distinctions, and our interest does not lie in differentiating between them. However, in general, the difference between consecutive levels is not known a priori. For this reason, at the training phase (including the training steps in the iterations of each K-fold CV), we did not merge these two classes and thus challenged the methods to select the best components for ordinal prediction when the differences in levels are extremely uneven. We merge them into a single level, denoted as 1, only to display and evaluate the confusion matrices. We select the penalty parameters, among a grid of proposed parameters, via K-fold Cross Validation, as detailed in Section \ref{penalty}, with $K=5$. The selected couple of parameters for foCCA are $(\lambda_1,\lambda_2)=(100,1000)$. For the analysis here presented, the dataset cleaned from outliers was used; hence, hereafter we do not consider robust estimators for the mean and the covariance operators of the quantity involved. Alternatively, one may use one of the robust estimation procedures present in the literature (see, e.g., \cite{bali11}). \\ The first row of Figure \ref{scores} displays the scatterplot of the first two functional scores of foCCA, alongside the marginal densities of these scores. Notably, the marginal densities of the first score distinctly separate the highest concentration levels, while the lowest four levels remain largely overlapped. The second component effectively distinguishes levels $\{0,1\}$ from levels $\{2,3,4\}$, i.e. ROIs without spots from those with spots and a low antigen concentration.\\ The method foCCA also provides a clear understanding of the difference, with respect to the most relevant features of the functional profiles, among the levels. Indeed, by observing the first ordinal components (Figure \ref{focca_comp}, right) combined with their respective functional components (Figure \ref{focca_comp}, left), we can understand, for every couple of consecutive levels, how much they are separated by the corresponding functional score. In addition, we can determine whether moving to the next level results in an increase or a decline of the functional score. In Figure \ref{deviations} it is clear that the first functional component gives a high weight to the final phase of the reaction (time $t\geq400$). The vector $\theta_1$, i.e. the first ordinal component takes high values in the last components, in particular between levels 5 and 6 (see the right plot in Figure \ref{focca_comp}), where $\theta_1$ has the maximum value. All these considerations are consistent with the fact that the highest levels are well separated in the 2D functional scores' space (Figure \ref{scores}, top) and the levels 5 and 6 are here the most distant. On the contrary, the second score seems to separate the lowest levels (spot vs. non-spot); this observation is coherent with the information given by the right plot in Figure \ref{focca_comp}, where the highest component of $\theta_2$ is 1:2, i.e. the increment on the ordinal score passing from level 1 to level 2 (see Section \ref{theta} to interpret the ordinal component). Lastly, in Figure \ref{focca_comp}, one may notice that the second functional component gives a negative weight in the initial slot (approximately $t<60$): this first phase, in practice, is fundamental to discriminate between levels 1 and 2, since the initial light intensity variation is due to the presence of the spot and the lower the concentration of antigen (null at level 2) the higher the power of this variation. In conclusion, foCCA is able to provide further analysis of the levels paired with the functional profiles and, in our case study, it is consistent with the biological knowledge. 

\subsection{Validation}
\begin{figure}[H]
    \centering   
    \includegraphics[scale=0.5]{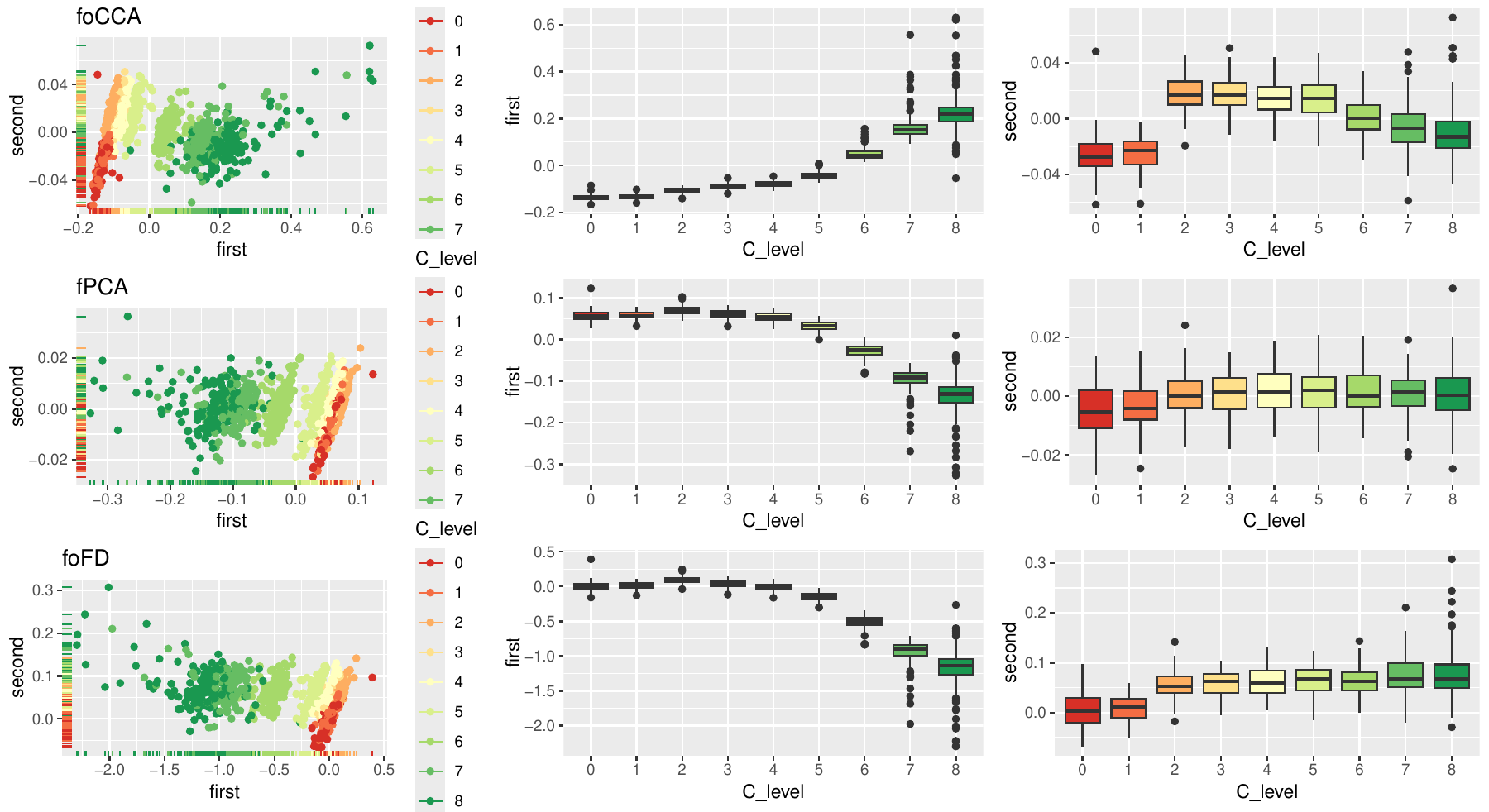}
    \caption{A comparison of the three methods. First column: the scatterplots of the first two scores. Second column: boxplots of the first score. Third column: boxplots of the second score.}
    \label{scores}
\end{figure}

\begin{figure}[H]
\centering\includegraphics[scale=0.55]{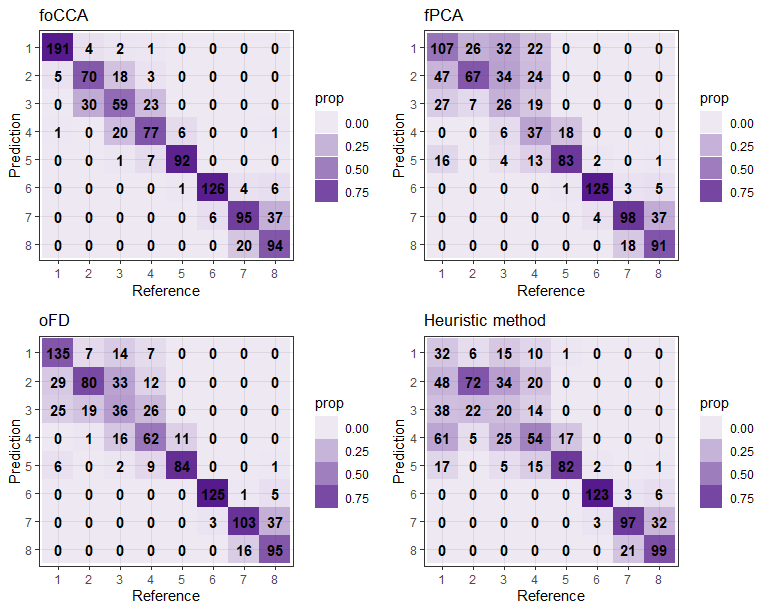}
    \caption{Comparison of the K-fold CV confusion matrices, with $K=5$}
    \label{confusion_mats}
\end{figure}

We now validate our method by comparing the dimensionality reduction obtained through foCCA with the ones obtained with fPCA, foFD, focusing on the ability to separate the consecutive levels of concentration. Moreover, we will compare the nearest centroid classifiers built on the first two components identified by foCCA, fPCA and foFD. For the sake of completeness, we shall also compare the results with a scalar competitor, built upon considering as score the last value of the log-ratio. The latter is a heuristic alternative to the first three methods, since, at a first sight, it can seem sufficient to predict the level of concentration. In order to have a fair comparison among foCCA, fPCA and foFD, we select the penalty parameters at the best of their ordinal classification power (analogously as in Section \ref{penalty}). The parameter selected for fPCA is $\lambda=10^6$, while for foFD is $\lambda=10^8$. Figure \ref{scores} shows that, as foCCA, fPCA and foFD provide a first score that distinctly separate the highest concentration levels. However, the second component of fPCA, contrarily to foCCA, does not exhibit clear separation between any pair of successive levels, while the second component of foFD separates the consecutive levels 1:2, but in a much weaker way than foCCA. \\
 By observing the confusion matrices in Figure \ref{confusion_mats}, it is clear that only foCCA has a high accuracy in the low levels of concentration. Moreover, fPCA, foFD and, even more, the heuristic method, show frequent misclassifications of 2 or more levels, especially when the target level is low, while in foCCA it happens much more rarely.\\A possible interpretation of these results can be given. Indeed, the functional direction along which the level 1 (i.e. 0 and 1) and the levels higher than 1 are most separated do not present a high global variance. For this reason, fPCA does not identify this direction as relevant. Moreover, the levels are not intrinsically equidistant, hence foFD tends to find the functional directions that maximize the separation among the most different successive levels. Contrarily, foCCA, through the use of vector $\theta$, naturally handles the varying distances of consecutive levels, approaching or moving away from levels to maximize correlation with the functional component. Lastly, the heuristic method, which would provide a fast and simple alternative to the functional methodologies, suggests that a scalar index does not contain enough information to distinguish all the levels. In conclusion, foCCA outperforms all the tested competitors in predicting the level of concentration of antigen, thanks to the flexibility gained with the ordinal canonical component $\theta$, and the targeting of the ordinal variable in its objective functional.

\begin{minipage}[t]{0.4\textwidth}
\begin{figure}[H]
    \centering
    \includegraphics[scale=0.42]{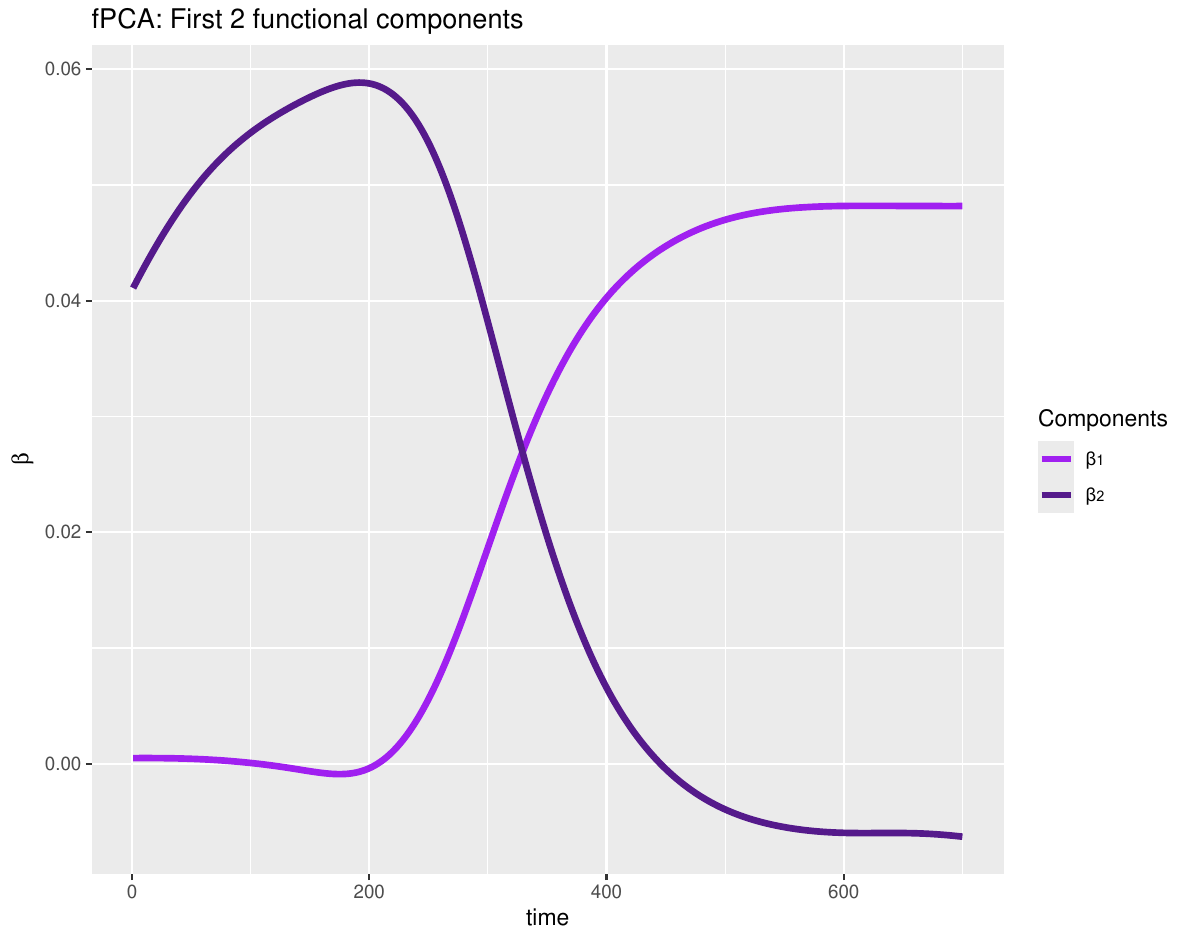}
    \caption{The first two components of fPCA}
    \label{pca_comp}
\end{figure}
\end{minipage}%
\hspace{1cm}
\begin{minipage}[t]{0.4\textwidth}
\begin{figure}[H]
    \centering
    \includegraphics[scale=0.42]{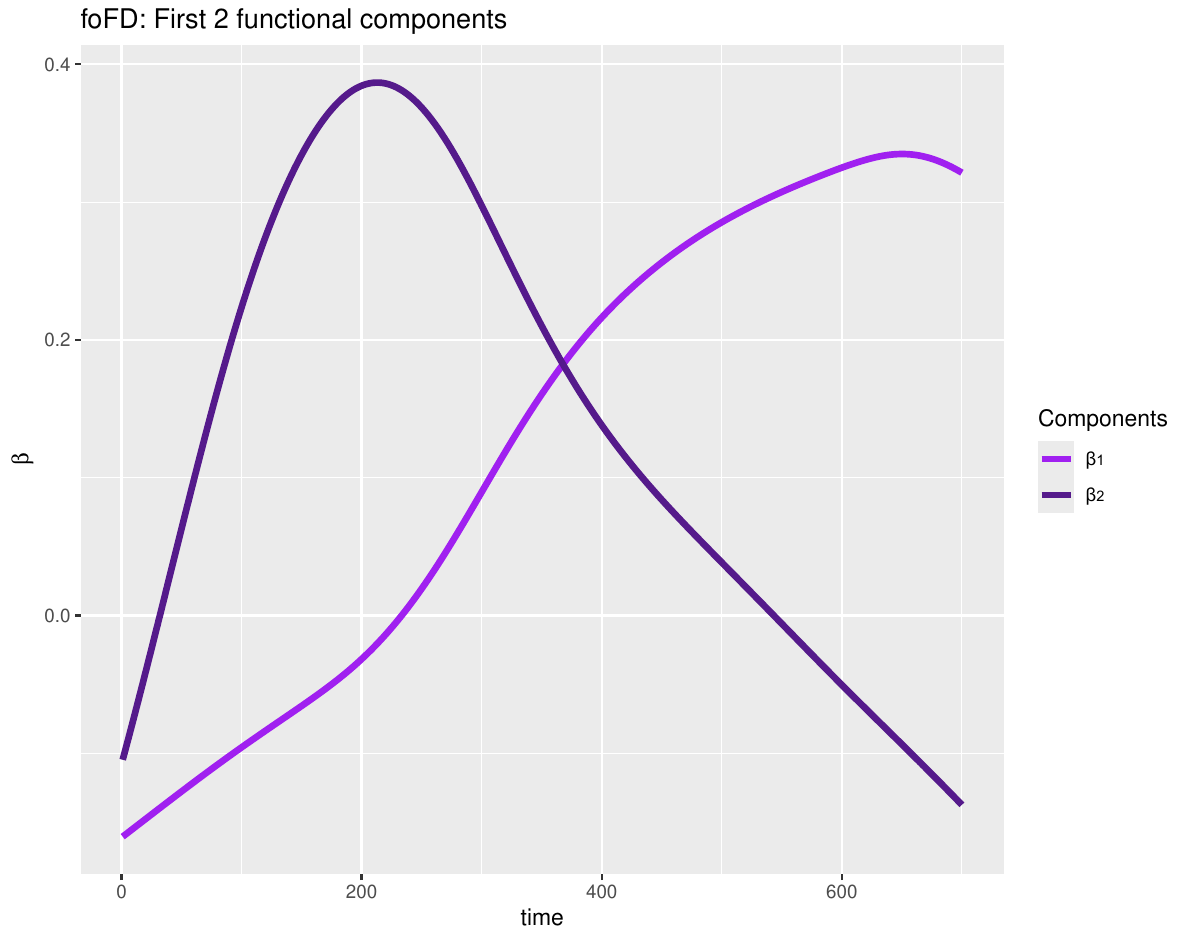}
    \caption{The first two components of foFD}
    \label{foFD_comp}
\end{figure}
\end{minipage}

\begin{minipage}{0.2\textwidth}
\begin{table}[H]
\centering
\begin{tabular}{@{}ll@{}}
\toprule
Method & Accuracy \\
\midrule
foCCA& 0.804\\
fPCA& 0.634 \\
foFD& 0.72 \\
heuristic& 0.579\\
\bottomrule
\end{tabular}
\caption{Accuracy}
\label{table}
\end{table}
\end{minipage}%
\begin{minipage}{0.8\textwidth}
\begin{figure}[H]
    \centering
    \includegraphics[scale=1.2]{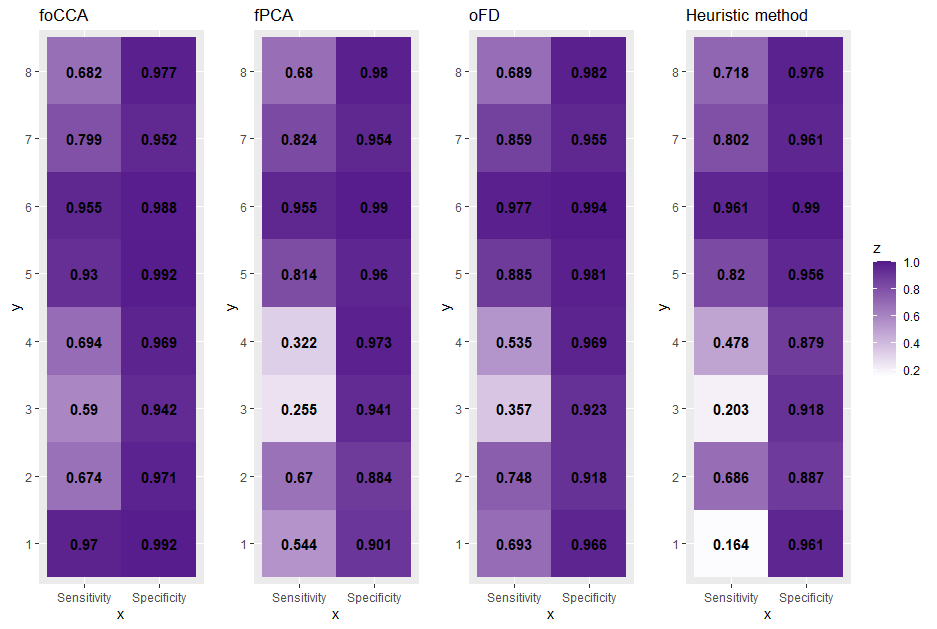}
    \caption{Comparison of sensitivity and specificity}
    \label{fig:enter-label}
\end{figure}
\end{minipage}


\section{Conclusions}
\label{sec:conc}
The work presented is inspired by a case study, when we want to maximize the ability to predict the antigen concentration level of each ROI, based on an optical signal. At this aim, we proposed a novel methodology to predict ordinal variables based functional data, foCCA. Through an extensive simulation study, we showed that foCCA performs significantly better, from an ordinal predictive point of view, than the competitors ( i.e. fPCA and foFD) in two realistic scenarios for any level of severity of both. In the case study, foCCA gives better ordinal predictive results than competitors, both functional (fPCA and foFD) and scalar (heuristic method) and, additionally, it provides signal-based distances between the consecutive levels, which in principle are not homogeneous. Moreover, the results are consistent with the physical knowledge about the ROIs. Our approach is particularly suitable when the final aim is ordinal prediction based on high dimensional data or when the signal-based differences between consecutive levels are considered of particular interest. As future improvement, the foCCA can be made able to predict in real-time the ordinal level. Indeed, in process control an early predicted level can be fundamental to take real-time actions, e.g. with the aim in view of building a digital twin.

\section*{Acknowledgements}
The authors gratefully thank Johanna Hutterer (University of T\"ubingen) for the interesting discussions about the biophysical consistency of the results and her contribution in the editing of the article. This research has received funding by the European Commission under the “HORIZON-CL4-2021-DIGITALEMERGING-01 project BioProS - Biointelligent Production Sensor to Measure Viral Activity” (grant agreement no. 101070120), 2022-2026”. GP, FN, LD, and AM acknowledge the initiative “Dipartimento di Eccellenza 2023–2027”, MUR, Italy, Dipartimento di Matematica, Politecnico di Milano. LD acknowledges his membership to INdAM GNCS - Gruppo Nazionale per il Calcolo Scientifico (National Group for Scientific Computing), Italy.





\end{document}